\documentclass[usenatbib,useAMS]{mn2e}
\usepackage{times}
\usepackage{epsfig}
\usepackage{amsmath}
\usepackage[usenames,dvips]{color}
\usepackage{subfigure}
\usepackage{rotating}
\usepackage{graphicx}
\usepackage{lscape}
\usepackage{pdflscape}
\usepackage{subfigure}
\usepackage{color}
\usepackage{upgreek}
\usepackage{float}

\title[The advantages of using a Lucky Imaging camera for observations of microlensing events]{The
advantages of using a Lucky Imaging camera for observations of microlensing events}

\author[Sedighe Sajadian, Sohrab Rahvar, Martin Dominik and Markus Hundertmark]
{Sedighe Sajadian$^{1}$ , Sohrab Rahvar$^{1}$, Martin Dominik$^{2}$ and Markus Hundertmark$^{3}$  \\
$^1$ Department of Physics, Sharif University of Technology, P.O. Box 11155-9161, Tehran, Iran \\
$^2$ SUPA, School of Physics \& Astronomy, University of St Andrews,
North Haugh, St Andrews, KY16 9SS, UK\\
$^{3}$Niels Bohr Institute \& Centre for Star and Planet Formation,
University of Copenhagen {\O}ster Voldgade 5, 1350 - Copenhagen,
Denmark}

\begin{document}


\maketitle
\begin{abstract}
In this work, we study the advantages of using a Lucky Imaging
camera for the observations of potential planetary microlensing
events. Our aim is to reduce the blending effect and enhance
exoplanet signals in binary lensing systems composed of an exoplanet
and the corresponding parent star. We simulate planetary
microlensing light curves based on present microlensing surveys and
follow-up telescopes where one of them is equipped with a Lucky
imaging camera. This camera is used at the Danish $1.54$-m follow-up
telescope. Using a specific observational strategy, For an
Earth-mass planet in the resonance regime, where the detection
probability in crowded-fields is smaller, lucky imaging observations
improve the detection efficiency which reaches $2$ per cent. Given
the difficulty of detecting the signal of an Earth-mass planet in
crowded-field imaging even in the resonance regime with conventional
cameras, we show that Lucky Imaging can substantially improve the
detection efficiency.
\end{abstract}

\section{Introduction}
Gravitational microlensing refers to the bending of light by the
gravitational potential of a stellar object
\cite{Einstein36,Paczynski86}. In this type of lensing, images from
the source can not be resolved by a typical ground based telescope
\cite{ChangRefesdal}. Mao and Paczy\'nski (1991) proposed using this
phenomenon for detecting planets, where planet and parent star
compose a binary system. This method has also been discussed earlier
by Liebes in 1964. In this type of microlensing events, a planet
produces anomalies in the light curve through the perturbation in
the gravitational potential of the primary lens \cite{gl92}. This
perturbation can have weak and strong features on the light curve of
a microlensing event, depending on the geometrical configuration of
the source star with respect to the caustic lines in the lens plane.
The observability of this anomaly is mostly sensitive to the planet
distance to the parent star on the lens plane, so-called 'lensing
zone'. This zone ranges in $[0.6,1.6] \mathrm{\theta_{E}}$
\cite{gl92,ben96,gs98} where $\mathrm{\theta_{E}}$ is the angular
Einstein radius and corresponds to the ring shape image when the
observer, source and lens are completely aligned. Moreover, in
close/wide high-magnification microlensing events when the source
star crosses the caustic lines close to the lens or the companion
planet, the planetary signals can be detected \cite{shin}.

The detectability of an exoplanet depends on the parameters of the
lensing system. These parameters are (i) planet to the host star
mass ratio, (ii) the angular separation, (iii) projected trajectory
of the source star on the lens plane and (iv) the size of the source
star. The advantage of planet detection by microlensing is that this
method enables exploring planets of Earth mass or even lower mass
planets \cite{Beaulieu,muraki,Dominik2008,Paczynski96}. While
Earth-mass planets have also been detected by the Kepler space
telescope  \footnote{http://kepler.nasa.gov/} and by the eclipsing
method (e.g. Quintana et al. 2014), the advantage of using
gravitational microlensing is to detect planets beyond the snow line
of their parent stars and explore planets orbiting low-mass stars
\cite{sumi1,bat}.

One of the problems in microlensing observations is the blending
effect of the source star. This effect refers to the contribution of
light from dense stellar fields within the Point Spread Function
(PSF) of the source star towards the Galactic bulge. The result is
imposing uncertainties on the calculation of the lens parameters.
One of the approaches for decreasing the blending effect is using
space-based telescopes for microlensing observations \cite{ben02}.
The other low-costs method for improving the angular resolution of
images from ground-based telescopes is using active adaptive optics
or Lucky Imaging cameras.

The Lucky Imaging or Lucky exposure technique was first proposed in
the 1960s to decrease the effect of atmospheric turbulence on
astronomical images (e.g. Hufnagel \& Stanley 1964). In this method,
a series of images is taken with exposure times shorter than the
time scale of the atmospheric turbulence. Then, the best images with
flatter wavefront are selected, re-centered and combined to get a
high resolution image. In order to quantify the quality of images,
we can use the Strehl ratio \cite{strehl} which is the ratio of
observed peak intensity to the theoretical peak intensity of a
perfect image at the diffraction limit (Baldwin et. al. 2001). The
first numerical calculation of Lucky Imaging in order to have a
Lucky short-exposure image was done by Fried (1978). The next
development was carried out by a group in Cambridge operating a
Lucky Imaging camera on the Nordic Optical Telescope (NOT) with
$2.56$ m aperture and obtained diffraction-limited images at wave
lengths of $810$ nm \cite{bal01,la06}. This group could improve the
Strehl ratio in visible bands on a large telescope by combining this
method with the adaptive optics method \cite{law08}. They could make
images with high resolutions from Mars and the other planets
\cite{dan00,cec07}.

We simulate the observing strategy with the Lucky Imaging camera
used for follow-up microlensing observations at the Danish $1.54$-m
telescope operated by the Microlensing Network for the Detection of
Small Terrestrial Exoplanets (MINDSTEp) consortium
\footnote{http://www.mindstep-science.org/}. This telescope follows
the observing strategy using a Lucky Imaging camera and EMCCDs
(Dominik et al. 2010). We introduce an optimum method for analyzing
light curves being observed both with Lucky Imaging and normal
cameras. The main goals of this strategy are (i) decreasing the
blending effect and enhancing signals from the anomalies on the
light curve by selecting only the high-quality images and (ii)
increasing the number of data points by considering all the photons
gathered by the telescope(s) (providing high speed photometry
enables us to correct instrumental effects and rebin and unrebin
datasets). Finally, we investigate the improvement on the planet
detection efficiency considering improvement of the blending effect
with using the Lucky Imaging camera.

The outline of this paper is as follows: in section (\ref{sec2}) we
investigate the blending effect on the detectability of exoplanet
microlensing events. In the next section, we review the Lucky
Imaging technique and study the best observational strategy for
detecting microlensing events via a Lucky Imaging camera. The effect
of using a Lucky camera on the planetary detection efficiency will
be discussed in section (\ref{detection}). Conclusions and summary
are given in section (\ref{con}).

\section{The effect of Blending on the detectability of exoplanet
microlensing signals} \label{sec2} The blending effect on the planet
detection efficiency as a function of the source radius was
quantitatively investigated by Vermaak (2000). Also, Gaudi \&
Sackett (2000a) introduced an algorithm to calculate the planet
detection efficiency for the observed microlensing light curves
which contain the blending effect. They noticed that it is crucial
the accurate determination of the blended light fraction for the
accurate determination of the efficiency of individual events. In
this part our aim is to show qualitatively the effect of blending on
the planet signatures.

The intrinsic magnification of a source star is suppressed by
including the blending of the background stars. These stars
contribute inside the area of Point Spread Function (PSF) of the
source star. Then the observed magnification $\mathrm{A_{obs}}$ as
the ratio of the observable flux to the baseline flux is related to
the intrinsic magnification of source star, $\mathrm{A_{\star}}$ by:
\begin{eqnarray}
A_{obs} = (1-b) + b A_{\star},
\end{eqnarray}
where $b$ is the blending parameter which is the ratio of the source
flux to the total baseline flux, without lensing effect:
\begin{eqnarray}\label{blendingP}
b=\frac{F_{\star}}{F_{\star}+F_{bg}},
\end{eqnarray}
where $\mathrm{F_{\star}}$ is the intrinsic flux of the source star
and $\mathrm{F_{bg}}$ is the baseline flux due to background stars
within the source PSF. For the case of $\mathrm{b=1}$ there is no
blending and $\mathrm{b=0}$ is the maximum blending. From the
microlensing observations, determining the intrinsic source and
baseline fluxes and as a result the blending parameter from light
curves is difficult and there is uncertainty in this parameter.
Hence, the blending effect can not always be removed from light
curves. On the other hand the blending of source star by increase
the level of background makes it hard to confirm the planetary
signals.

\begin{figure}
\begin{center}
\psfig{file=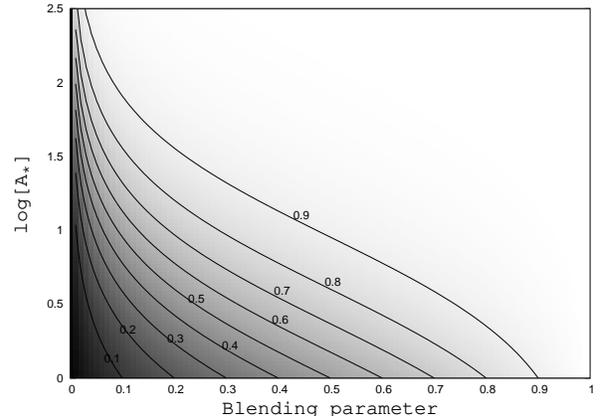,angle=270,width=8.0cm,clip=} \caption{Relative
observed signal to the relative intrinsic signal, e.g. due to an
exoplanet, in a microlensing light curve, i.e. equation (\ref{y}),
as a function of the intrinsic magnification of the source star
$\mathrm{A_{\star}}$ and the blending parameter $b$.} \label{fig1}
\end{center}
\end{figure}
In order to quantify the effect of blending on an exoplanet signal
in the light curve, let us assume the effect of companion planet in
the binary lens is a perturbation on the Paczy\'nski's single lens
light curve. Then, a perturbation $\mathrm{\delta A_{\star}}$ is
related to the observed magnification $\mathrm{\delta A_{obs}}$, as
follows
\begin{eqnarray}
(\delta A/A)_{obs} =  ({\delta A}/{A})_{\star} \frac{
A_{\star}}{A_{\star} + b^{-1} -1 }, \label{eq2}
\end{eqnarray}
where we can call $\mathrm{\delta A/A}$ as the relative contrast and
for two extreme limits of $\mathrm{b \rightarrow  0}$, we get no
signal from the exoplanet and for the case that $\mathrm{b=1}$, we
get the maximum signal.

\begin{figure}
\begin{center}
\psfig{file=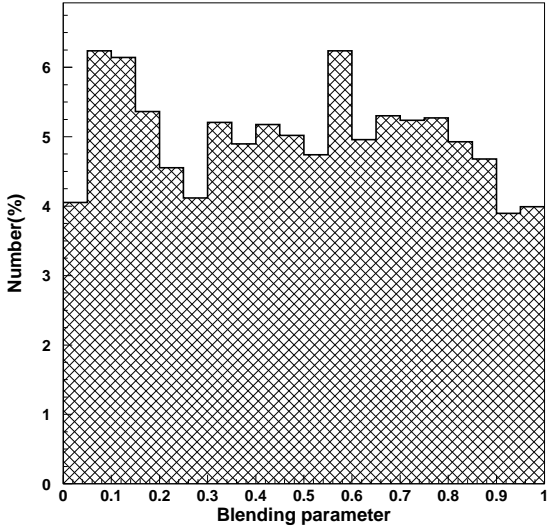,angle=0,width=8.0cm,clip=} \caption{The
distribution of blending parameter towards the Galactic bulge for
$3560$ OGLE-III microlensing events (Wyrzykowski et al.
2014).}\label{fig2}
\end{center}
\end{figure}
We can define the relative observed signal to the intrinsic signal
of source star by
\begin{eqnarray}
Y = \frac{(\delta A/A)_{obs}}{({\delta A}/{A})_{\star}}, \label{y}
\end{eqnarray}
here the denominator of this expression (i.e. $\mathrm{({\delta
A}/{A})_{\star}}$) depends on the parameters of the binary lens as
the distance of the planet from the parent star, the mass ratio and
relative position of the lenses and the source star as well as the
angular size of the source star. On the other hand, the numerator
(i.e. $\mathrm{({\delta A}/{A})_{obs}}$) depends on the signal to
noise ratio of the observed data in the light curve.

Figure (\ref{fig1}) represents the $\mathrm{Y}$ parameter as a
function of the intrinsic magnification of the source star and the
blending parameter. In order to get a strong signal from the planet
either the magnification of the source star should be high enough or
the blending parameter needs to be larger. The amount of blending
depends on the luminosity of the source star and density of the
field. The distribution of the blending parameter of microlensing
events for $3560$ events, observed by the OGLE collaboration towards
the Galactic bulge for the period of years 2001-2009 \cite{OGLE3} is
given in Figure (\ref{fig2}).

In what follows, we investigate the blending effect on the
detectability of an exoplanet signal in the flux. We perturb the
source flux as a result of the presence of an exoplanet. The
relative observed variation of flux relates to the relative
intrinsic variation of flux as:
\begin{eqnarray}
(\frac{\delta F}{F})_{obs} = b (\frac{\delta F}{F})_{\star}.
\end{eqnarray}
We can rewrite this equation in terms of magnitude as
$\mathrm{\Delta m_{obs} = b \Delta m_{\star}}$. Having larger $b$
(e.g. with the Lucky Imaging method that will be discussed later),
we can observe larger variations in the magnitude of the star. Let
us assume the magnification of simple microlensing light curve is
$\mathrm{A_{\star}}$ and any perturbation around it is
$\mathrm{\delta A_{\star}}$. Then, the overall flux taking into
account the background flux would be:
\begin{eqnarray}
F_{obs} = F_{\star}(A_{\star} + \delta A_{\star}) + F_{bg},
\end{eqnarray}
where using the definition of blending parameter which is brought in
the equation (\ref{blendingP}):
\begin{eqnarray}
F_{obs} = F_{\star}( A_{\star} + \delta A_{\star} + b^{-1} - 1).
\end{eqnarray}
The photometric error bar, due to error in the source flux as well
as other sources for the noises is given by
\begin{eqnarray}
\delta F_{obs}= \frac{h\nu}{t_{exp}}\sqrt{\frac{t_{exp}}{h\nu}
F_{\star}( A_{\star} + \delta A_{\star} + b^{-1} - 1)+C(t)^{2}},
\end{eqnarray}
where $\mathrm{\nu}$ is the frequency of the light,
$\mathrm{t_{exp}}$ is the exposure time and $C$ represents other
noise terms excluding noise in the source flux (e.g. readout noise,
dark noise, constant gain, extra noise from fast readouts for the
Lucky Imaging camera, etc). We note that $C$ generally depends on
time. This is because the blending effect in microlensing events is
usually estimated via difference imaging and is therefore affected
by the choice of reference frame. The reference frames for different
observational groups are chosen at different times. Since the
baseline flux changes with time, so changes $\mathrm{C}$
representing other noise terms except for the time-dependent
uncertainty in source flux. However, the photometric error depends
on the number of pixels which is not considered here.

The observability of a signal means that a perturbation in the
observed flux, $\mathrm{\Delta F}$, would be at least larger than
the photometric error bar, i.e. $\mathrm{\Delta F
> \delta F_{obs}}$. According to the amount of the perturbation in the
observed flux as $\mathrm{\Delta F = F_{\star}\delta A_{\star}}$,
the observability condition results in:
\begin{eqnarray}
N_{p}>\frac{1}{\delta A_{\star}^2} \left[A_{\star} +\delta A_{\star}
+b^{-1}-1+\frac{C(t)^2 h\nu}{F_{\star}t_{exp}}\right],
\end{eqnarray}
where $\mathrm{N_{p} = F_{\star} t_{exp}/(h\nu)}$ represents the
number of photons from the source star for a given exposure time.
Having smaller $b$ requires a larger number of photons. In the next
section, we will discuss the possibility of decreasing the blending
effect using the Lucky Imaging camera.
\begin{figure*}
\begin{center}
\psfig{file=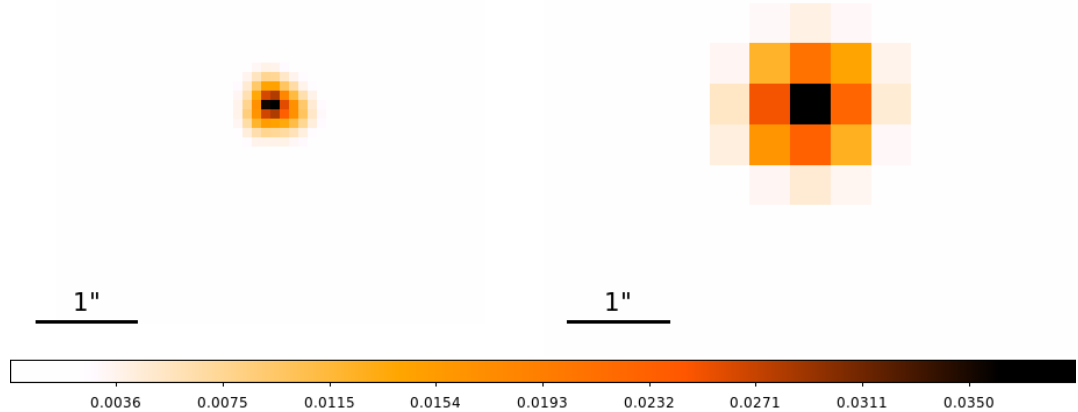,angle=90,width=0.8\textwidth,clip=0}\caption{
Lucky Imaging of an averaged PSF by the Danish telescope. The left
panel shows a diffraction-limited stellar PSF extracted from the
Lucky Imaging camera obtained from the DANDIA pipeline (Bramich
2008). The right panel represents the same image taken by the Danish
Faint Object Spectrograph and Camera (DFOSC). The image scales are
$0.09\mathrm{~arcsec/pixel}$ for the Lucky Imaging camera and
$0.39\mathrm{~arcsec/pixel}$ for DFOSC.} \label{fig4}
\end{center}
\end{figure*}

\section{Observations of planetary microlensing events with a Lucky Imaging camera}
\label{gl} In this section we investigate the characteristics of the
Lucky Imaging method and introduce a strategy for the observation of
planetary microlensing events with the Lucky Imaging camera. We
assume that the Lucky Imaging observation is carried out by the
Danish telescope which is one of the follow-up telescopes.

\subsection{The Lucky Imaging method}
One of the challenging problems of photometry and imaging in
astronomy is dealing with atmospheric seeing variations. The seeing
parameter increases with atmospheric turbulence where the wave front
of a distance source gets distorted by crossing different layers of
the atmosphere. Usually a telescope receives distorted wave fronts;
but occasionally we can find flat wave fronts. The probability of
finding such a wave front depends on the size of the wave front
patch arriving at the primary mirror of the telescope. Hence, we
expect that smaller telescopes are more likely to receive flat wave
fronts compared to larger telescopes in optical wavelengths. The
probability of a diffraction-limited image detected by a telescope
with an aperture size of $D$ in an instant of time while the wave
front crosses the atmospheric turbulence is given by \cite{fr78}:
\begin{eqnarray}\label{11}
P\simeq 5.6 \exp{(-0.1557\times{D^2/r_{0}^2})},
\end{eqnarray}
where $\mathrm{r_{0}}$ is the coherent length which is defined as
the diameter of an area over which the root mean square (rms) of
phase variation due to the atmospheric turbulence is equal to $1$
radian \cite{fr78}. This parameter defines the seeing of observation
equal to $\mathrm{\sim251\lambda/r_{0}}$ where $\mathrm{\lambda}$ is
the wave length in $\mu$m, $\mathrm{r_{0}}$ is in mm and seeing is
given in the unit of the arc second. At wave length of $\sim500$ nm, the
most common value of $\mathrm{r_{0}}$ is about $10-15$ cm, but in
good conditions the value of this parameter reaches to $30-40$ cm.
So the least amount of $\mathrm{D/r_{0}}$ for $1$-meter class
telescopes in visible band is about $\sim 3$. We can interpret
$\mathrm{D/r_{0}}$ as the ratio of "seeing" to "diffraction limit"
of a telescope:
\begin{eqnarray}
\frac{\theta_{seeing}}{\theta_{diff}} = \frac{D}{r_0},
\end{eqnarray}
where minimum discernable angular separation of two sources by
considering diffraction limit is equal to
$\mathrm{\theta_{diff}(rad)=1.22\lambda(cm)/D(cm)}$. The probability
function for taking a Lucky short exposure image  from equation
(\ref{11}) is a decreasing function versus $\mathrm{D/r_{0}}$ where
$\mathrm{D/r_{0}}$ usually ranges between $\mathrm{4 <D/r_{0}<7}$
\cite{Hec}.

In Lucky Imaging, we select a fraction of images to construct a
final image with a smaller Full Width at Half Maximum (FWHM) value
or in other words larger Strehl ratio \cite{law08}. However, just by
re-centering all taken images with the Lucky camera, we can already
decrease the FWHM (see Figure (4) of Law et al.2006). A decrease in
FWHM in a given seeing amount is quantified by the so-called
improvement factor (IF), defined as:
\begin{eqnarray}
IF= \frac{FWHM_{\star}}{FWHM_{lc}},
\end{eqnarray}
where $\mathrm{FWHM_{\star}}$ is the mean of FWHM of a typical
source star observed by a telescope without using the Lucky Imaging
camera for a given amount of seeing and $\mathrm{FWHM_{lc}}$ is the
FWHM of the final image produced by a Lucky Imaging camera
\cite{la06}. Hence, using Lucky Imaging the stellar PSF decreases
and can even reach the diffraction limit. Figure (\ref{fig4})
illustrates this point: the left panel represents a
diffraction-limited stellar PSF extracted from the Lucky Imaging
camera at the Danish $1.54$-m telescope (EMCCD) obtained from the
DANDIA pipeline \cite{Bramich2008}. The right panel shows the same
image taken by the Danish Faint Object Spectrograph and Camera
(DFOSC). The image scales are $\mathrm{0.09~arcsec/pixel}$ for the
Lucky Imaging camera and $\mathrm{0.39~arcsec/pixel}$ for DFOSC. The
PSF area for the image taken by Lucky Imaging (left panel) is about
$\mathrm{0.2~arcsec^2}$ and that for the image taken by DFOSC (right
panel) is $\mathrm{1.1~arcsec^2}$, where we define the PSF area by
$\Omega=\pi(FWHM/2)^{2}$. Consequently, this method reduces the PSF
area and blending by the background stars. By selecting only the
best images, the final image will have a smaller PSF area or higher
IF.

\begin{figure*}
\begin{center}
\psfig{file=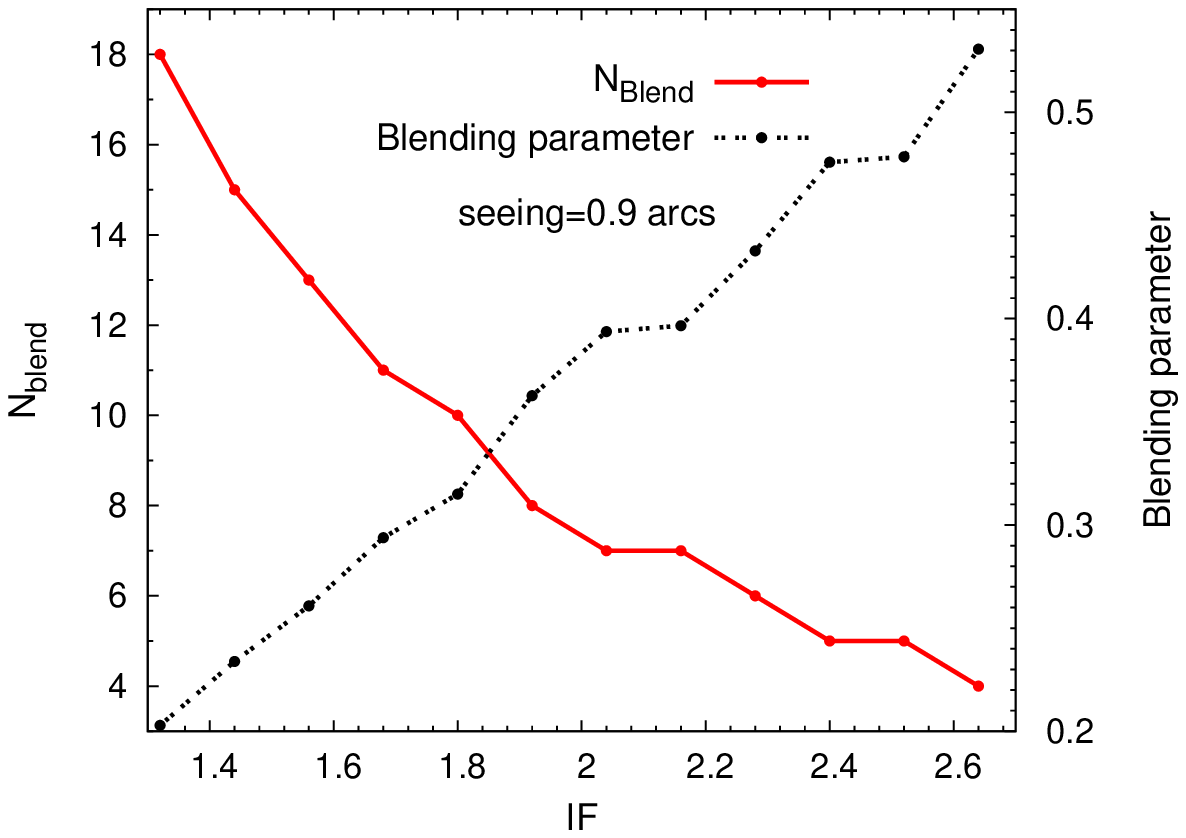,angle=0,width=8.cm,clip=}
\psfig{file=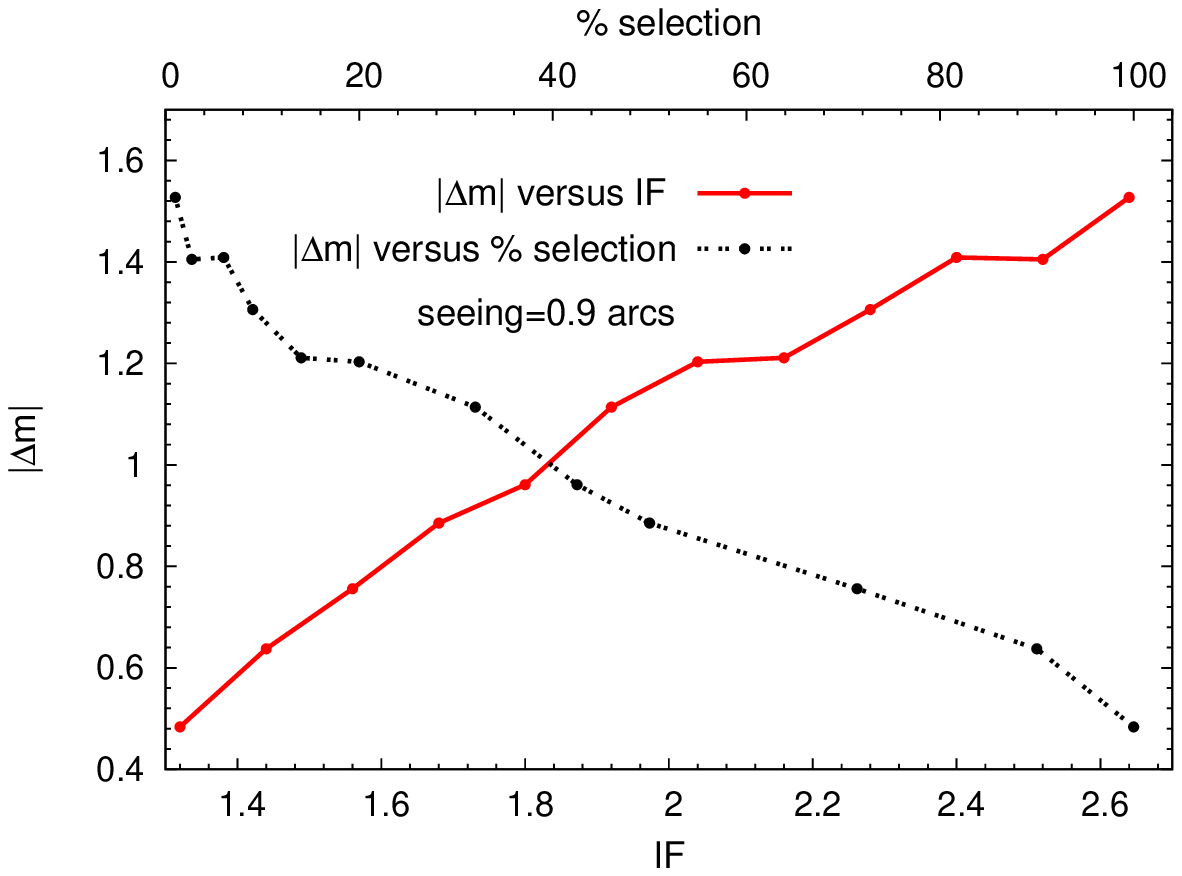,angle=0,width=8.cm,clip=} \caption{Left panel:
the number of blended stars whose lights enter the PSF of source
star (red solid line) and blending parameter (black dotted line) as
a function of IF. Right panel: variation of apparent magnitude of
the stellar PSF versus IF (red solid line) and percentage of the
selected frames (black dotted line). Here, we take the angular
distribution of stars in the Galactic bulge from the synthetic
Besan\c{c}on model (Robin et al. 2003) and set the apparent
magnitude of the source star is $\mathrm{m_{\star}=20.0}$ mag, the
initial amount of the blending parameter is $\mathrm{b=0.13}$ and
$\mathrm{seeing}=0.9~\mathrm{arcsec}$.} \label{fig3}
\end{center}
\end{figure*}
In order to find an optimum procedure of observation with the Lucky
Imaging method, we consider a highly-blended area of stars in the
Galactic bulge and determine the blending parameter as follows.
First we calculate the average number of blended stars by
integrating over the number density of the Galactic bulge stars
along the the line of sight and the boundary of integral within the
area of the source PSF, i.e. $\mathrm{<N_{blend}>=\int
\rho(D,\Omega_{\star})D^{2} dD\Omega}$, where $\mathrm{D}$ is the
distance from the observer position, $\mathrm{\Omega_{\star}}$ is
the angular position of the source star. We take the number density
of the Galactic bulge, $\mathrm{\rho}$, from the Besan\c{c}on model
\cite{dwek95,Besancon}. 
For these stars, we have the absolute
magnitude according to their mass, age and metallicity. Then, the
apparent magnitude of each blended star is calculated according to
the absolute magnitude, distance modulus and extinction. The
extinction map towards the Galactic bulge as a function of distance
is adapted from Gonzalez et al. (2012).

Figure (\ref{fig3}) shows the number of background stars which
contribute in the PSF of source star (red solid line) and the
blending parameter (black dotted line) versus IF. We note that
increasing IF decreases the PSF area of the source star for a given
amount of seeing. Also we show (in the left panel) the variation of
the apparent magnitude in that stellar PSF versus IF (red solid
line) and fraction of selected frames (black dotted line). Here, we
use the relation between percentage of selected frames and IF for
the fixed seeing amount (i.e. $\mathrm{0.9~arcsec}$) given by Law et
al. (2006). However, according to the Fried(1978)'s relation the
probability of getting a diffraction-limited image by the Danish
telescope at $\mathrm{R}$ band is even a little higher than that by
NOT at $\mathrm{I}$ band, if it would not be optics-limited.
Decreasing the PSF area with the Lucky Imaging method deblends the
apparent magnitude of the stellar PSF due to decreasing the number
of blended stars. In these figures, we set the apparent magnitude of
the source star to $\mathrm{m_{\star}=20.0}$ mag, the initial amount
of the blending is $\mathrm{b=0.13}$ and
$\mathrm{seeing=0.9~\mathrm{arcsec}}$. Noting that in reality seeing
is changing with time during the observation. Hence, we choose the
different fractions of images in different seeing amounts to have
constant PSF area throughout the observation.
\begin{figure}
\begin{center}
\psfig{file=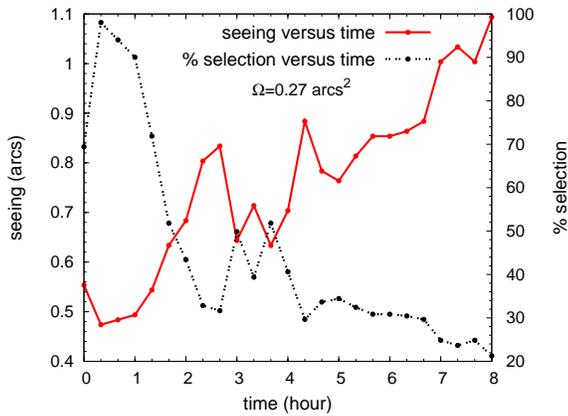,angle=0,width=8.cm,clip=}\caption{Seeing
variation versus time (red solid line) for every $20$ min and the
corresponding percentage of selected frames (black dashed line) to
get a constant PSF area of the source star, $\mathrm{{\Omega}=0.27
~arcsec^{2}}$.} \label{fig5}
\end{center}
\end{figure}

Lucky imaging observations reduce the blending and the PSF size by
selecting high quality images and the aim to (i) better constrain
the blending parameter, (ii) make a more reliable fit of the
parameters of the microlensing light curves and (iii) detect small
perturbations in the light curve, such as planetary signals (see
equation \ref{eq2}).

\subsection{An observational strategy for using a LI camera}
In what follows, we introduce a strategy for analyzing microlensing
light curves hypothetically being observed both by the Lucky Imaging
and normal cameras. Our aim is to optimize the quality of the light
curves, by means of having less blending of the source star, enough
cadence in the light curve and the least photometric uncertainties
for data points. In the simulation we keep the two crucial
parameters constant throughout the light; (i) we select the frames
from the Lucky imaging camera in such a way that the source PSF area
is a constant parameter and (ii) for each data point, we stack
enough images from the lucky camera to have constant photometric
accuracy throughout the light curve.

We note that in reality during the observation, the "seeing"
parameter changes with time and in order to have a constant PSF area
of the source star, we should have a time dependent procedure for
choosing percentage of the selected frames from each spool of
images. An example that demonstrate the variation of seeing versus
time is shown in Figure (\ref{fig5}). Here, the seeing parameter
(red solid line) is shown with the cadence of $20$ minutes. In our
simulation, the seeing amounts are chosen from a synthetic series of
seeing data points which were archived by the La Silla observatory
\footnote{https://www.eso.org/sci/facilities/lasilla/astclim/seeing.html}

Keeping a constant value for the PSF area of the source star (e.g.
$\mathrm{\Omega=0.27~arcsec^{2}}$), we calculate the percentage of
the selected frames corresponding to each value of seeing. Following
this strategy, during the microlensing event, blending parameter
which is specified by the PSF area of source star is set to be a
constant value. On the other hand, adapting the second condition, we
use an enough number of images to get a threshold value for the
photometric precision. This factor can be tuned by the effective
exposure time for each data point ($\mathrm{t_{exp}}$) to have
uniform error bars throughout the light curve, as suggested in
Dominik et al.$(2010)$. Since the magnification factor for each data
point is different, the signal to noise ratio is different at each
moment. According to the definition, the photometric accuracy is
given by:
\begin{eqnarray}
\frac{\sigma_{F}}{F}=\frac{\sigma_{N_{p}}}{N_{p}},
\label{limmit}
\end{eqnarray}
where $N_{p}$ is the total number of photons collected in PSF of the
source star and $\mathrm{\sigma_{N_{p}}=\sqrt{N_{p}}}$ is the
photon-count noise. To get uniform error bars for data points the
signal to noise ratio for each data point should reach to a
threshold amount where signal is defined as the total number of
photons:
\begin{eqnarray}
N_{p}&=&[(A_{\star}-1)10^{-0.4m_{\star}}+\Omega10^{-0.4 \mu_{sky}}\nonumber\\
&+&\sum_{i=1}^{N_{blend}}10^{-0.4 m_{i}}]10^{0.4m_{zp}}\times t_{exp},
\end{eqnarray}
where $\mathrm{A_{\star}}$ is the magnification factor of the source
star, $\mathrm{m_{\star}}$ is the apparent magnitude of source star,
$\mathrm{m_{i}}$ is the apparent magnitude of $i$th background star
within the PSF of source star, $\mathrm{N_{blend}}$ is the total
number of blended stars, $\mathrm{\mu_{sky}}$ is sky brightness in
$\mathrm{mag/arcsec^{2}}$ unit and $\mathrm{m_{zp}}$ is the zero
point magnitude which is equivalent to a photon collected over the
unit of the area of the telescope per unit of time. According to the
definition of magnitude, $\mathrm{m_{v}^{zp}}$ is given by:
\begin{eqnarray}
m_{v}^{zp}= -2.5 \log (\frac{h\nu}{\pi (D/2)^{2}F_{0}}),
\end{eqnarray}
where $\mathrm{F_{0}}$ is the absolute flux corresponding to the
magnitude of zero. \cite{bessel}. We keep the signal to noise ratio
of $500$ for lucky imaging data points throughout the light curve.
Note that the total observing time in terms of the effective
exposure time will be: $\mathrm{T_{obs}= t_{exp} / f}$, where
$\mathrm{f}$ is the fraction of selected frames. We note that
$\mathrm{T_{obs}}$ is longer than $t_{exp}$.
\begin{figure}
\begin{center}
\psfig{file=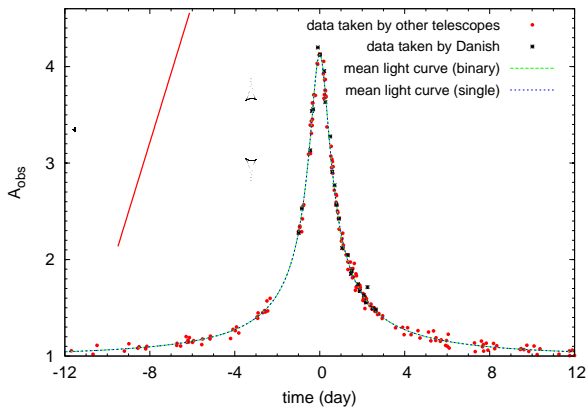,angle=270,width=8.cm,clip=} \caption{A
simulated planetary microlensing light curve. Its theoretical light
curve is shown with green dashed line. The simulated data points
over this light curve hypothetically taken by the Danish telescope
and other (survey and follow-up) telescopes are represented by black
stars and red circles respectively. The parameters used to make this
light curve are $\mathrm{u_{0}=-0.04}$, $\mathrm{t_{E}=12 days}$,
$\mathrm{t_{0}=0}$, $\mathrm{\rho_{\star}=0.003}$,
$\mathrm{q=3e-5}$, $\mathrm{d(R_{E})=0.9}$,
$\mathrm{\theta=25^{\circ}}$, $\mathrm{m_{\star}=20.2~mag}$ and
$\mathrm{b=0.13}$. } \label{fig6}
\end{center}
\end{figure}

In reality a Lucky Imaging camera needs extra exposure times due to
(i) the extra noise at the electron multiplication stage (the
cascade can trigger unwanted extra electrons), (ii) extra noise by
distributing the flux over many more pixels in case of not so
optimal seeing or the worst seeing bins and (iii) extra noise due to
the shift-and-add. The latter aspects are somewhat related, since
the signal to noise ratio generally depends on the number of pixels
used. The more we shift images, the more the different uncertainties
in pixel-to-pixel sensitivity i.e. in the flatfield, sky, etc. All
these aspects need more exposure times. The signal to noise ratio
for the Lucky imaging camera is lower by a factor of
$\mathrm{\sqrt{2}}$ as compared to regular CCDs
\cite{skottfelt2014}. This penalty can be avoided if the EMCCD is
operated in photon counting mode.

Here, we simulate the planetary microlensing light curve shown in
Figure (\ref{fig6}), being observed by a group of telescopes where
one of them is the Danish $1.54$-m telescope. The parameters used to
generate this light curve are $\mathrm{u_{0}=-0.04}$,
$\mathrm{t_{E}=12 ~days}$, $\mathrm{t_{0}=0}$,
$\mathrm{\rho_{\star}=0.003}$, $\mathrm{q=3\times 10^{-5}}$,
$\mathrm{d(R_{E})=0.9}$, $\mathrm{\theta=25^{\circ}}$,
$\mathrm{m_{\star}=20.2~mag}$ and $\mathrm{b=0.13}$ where
$\mathrm{\theta}$ is the angle between the projected source
trajectory and the binary axis and $\mathrm{\rho_{\star}}$ is the
projected source radius normalized to the Einstein radius.

To produce synthetic data points over this light curve, we assume
this light curve is being observed by survey and follow-up
telescopes while follow-up observations are started after the
magnification factor reaches to a given threshold (e.g.
$\mathrm{\frac{3}{\sqrt{5}}}$). In order to make a realistic light
curve in the simulation, cadences between the data points and the
photometric uncertainties are taken from the archived microlensing
model light curves along with data collected by the Optical
Gravitational Lensing Experiment (OGLE), Microlensing Observations
in Astrophysics (MOA), MiNDSTEp, follow-up observations of
microlensing events with a robotic network of telescopes
(ROBONet-II), Microlensing Follow-Up Network (MicroFUN) and Probing
Lensing Anomalies Network (PLANET)
\footnote{http://ogle.astrouw.edu.pl/ogle4/ews/ews.html}. We adapt
$20$ archived microlensing events and obtain the sampling time
between consecutive data points from the observed data over the
light curves as well as the photometric accuracy of data points.

We separate the simulated data points into two categories of being
$\mathrm{A_{obs}\leq3/\sqrt{5}}$ where only the survey telescopes
with the cadence and photometry error of data associated with the
surveys are chosen and $\mathrm{A_{obs}\geq3/\sqrt{5}}$ after
alerting the event where the specification of the surveys and
follow-up telescopes are adapted. In order to make lucky imaging
observation with the Danish telescope we label archived data points
in reality taken by this telescope which is illustrated by black
stars in Figure (\ref{fig6}).  Here we use the adaptive contouring
method to calculate the magnification of the source star in the
binary lensing \cite{Dominik2007}. The simulated data points are
shifted with respect to the mean light curve according to their
photometric uncertainties by a Gaussian function.

Figure (\ref{fig6}) as a sample represents the data points over the
simulated light curve. The data points by the Danish $1.54$-m
telescope are in black stars and other surveys and follow-up
telescopes are in red points. The inset represents the caustic
curves (black solid curves) and the source trajectory with respect
to the lens position projected on the lens plane (red solid line).
Also the microlensing light curve with the planetary lens and simple
Paczy\'nski microlensing light curve are shown by green dashed and
blue dotted lines, respectively.

In the next step, we regenerate this light curve by replacing data
from the conventional camera in the Danish telescope with data taken
by the Lucky camera. For simulating these data, the important
quantity is the time interval between two consecutive data points.
Since the Danish telescope simultaneously observes several
microlensing targets during each night, so there is a time interval
i.e. cadence, between data points of each light curve. There is an
experimental relation between cadence $\mathrm{\tau}$ and the
observed magnification factor \cite{Dominik10}:
\begin{eqnarray}
\tau= 90 \sqrt{\frac{3/\sqrt{5}}{A_{obs}}}~~\rm{ min}.
\end{eqnarray}
The crucial point regarding the sampling interval is that we set it
to $\mathrm{\sim 10}$ minutes once at least three data points
deviate from the simple microlensing light curve by more than
$\mathrm{2 \sigma}$. Once the anomaly is covered, we return to the
normal sampling rate. In addition to the above cadence, we should
consider the overhead time in between such as slew time and
operational losses \cite{Dominik10}. The operational loss could be
due to the weather, technical and operational issues. The average
values of the overhead time in between two consecutive data points
and the lost time for each data point taken by the Danish telescope
are $\mathrm{\sim1.57}$ min and $\mathrm{\sim 1.45}$ min, respectively.
Considering Poisson uncertainties for these time scales, they are
added to these two time scales.
\begin{figure*}
\begin{center}
\psfig{file=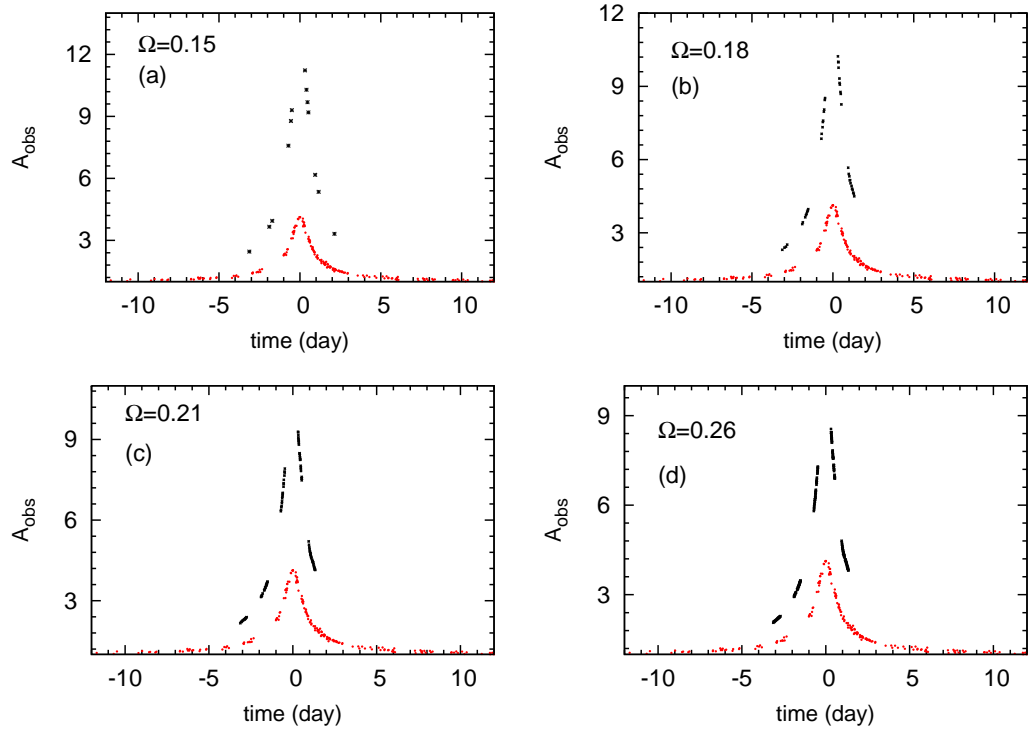,angle=270,width=0.8\textwidth,clip=} \caption{
Each panel shows the simulated microlensing light curve shown in
Figure (\ref{fig6}) which is modified by replacing data from the
conventional camera on the Danish telescope with synthetic data
hypothetically taken by Lucky Imaging (black stars), the rest of
data points hypothetically taken by usual cameras (red circles). The
light curves are plotted for four different values of
$\mathrm{\Omega}$.} \label{fig7}
\end{center}
\end{figure*}
\begin{figure*}
\begin{center}
\psfig{file=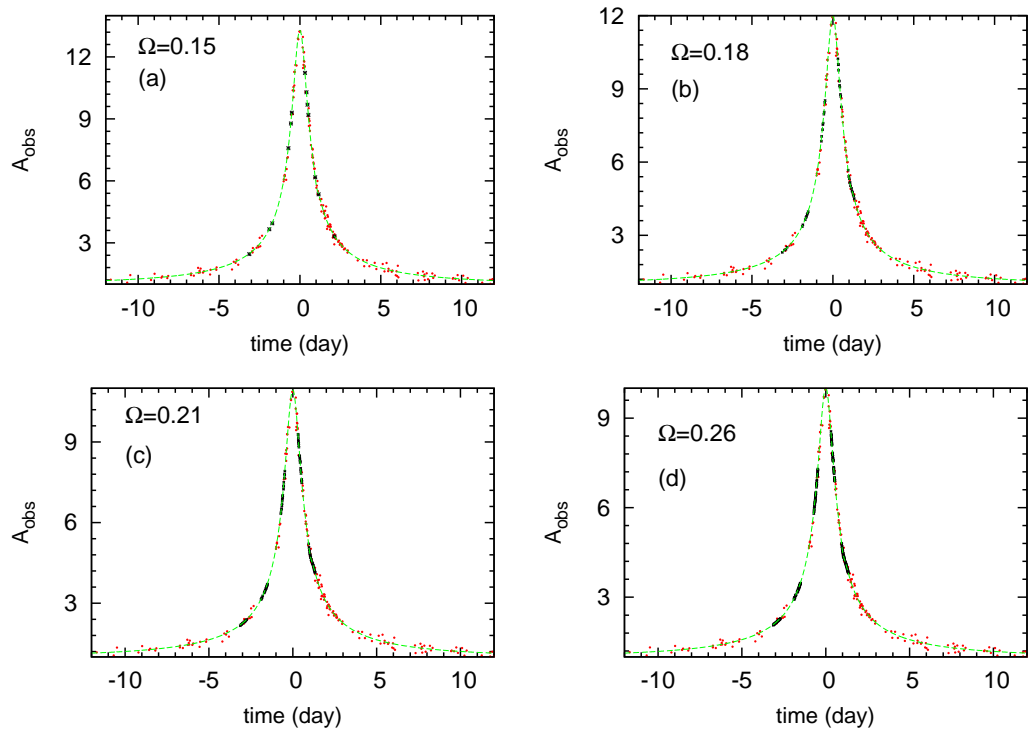,angle=270,width=0.8\textwidth,clip=0}
\caption{Data points hypothetically taken by normal cameras (red
circles) in light curves of Figure (\ref{fig7}) are shifted to the
lower blending data points taken by Lucky Imaging. This shifting is
done for various $\mathrm{\Omega}$s using $\mathrm{\alpha}$
parameters.} \label{fig8}
\end{center}
\end{figure*}
\begin{figure*}
\begin{center}
\psfig{file=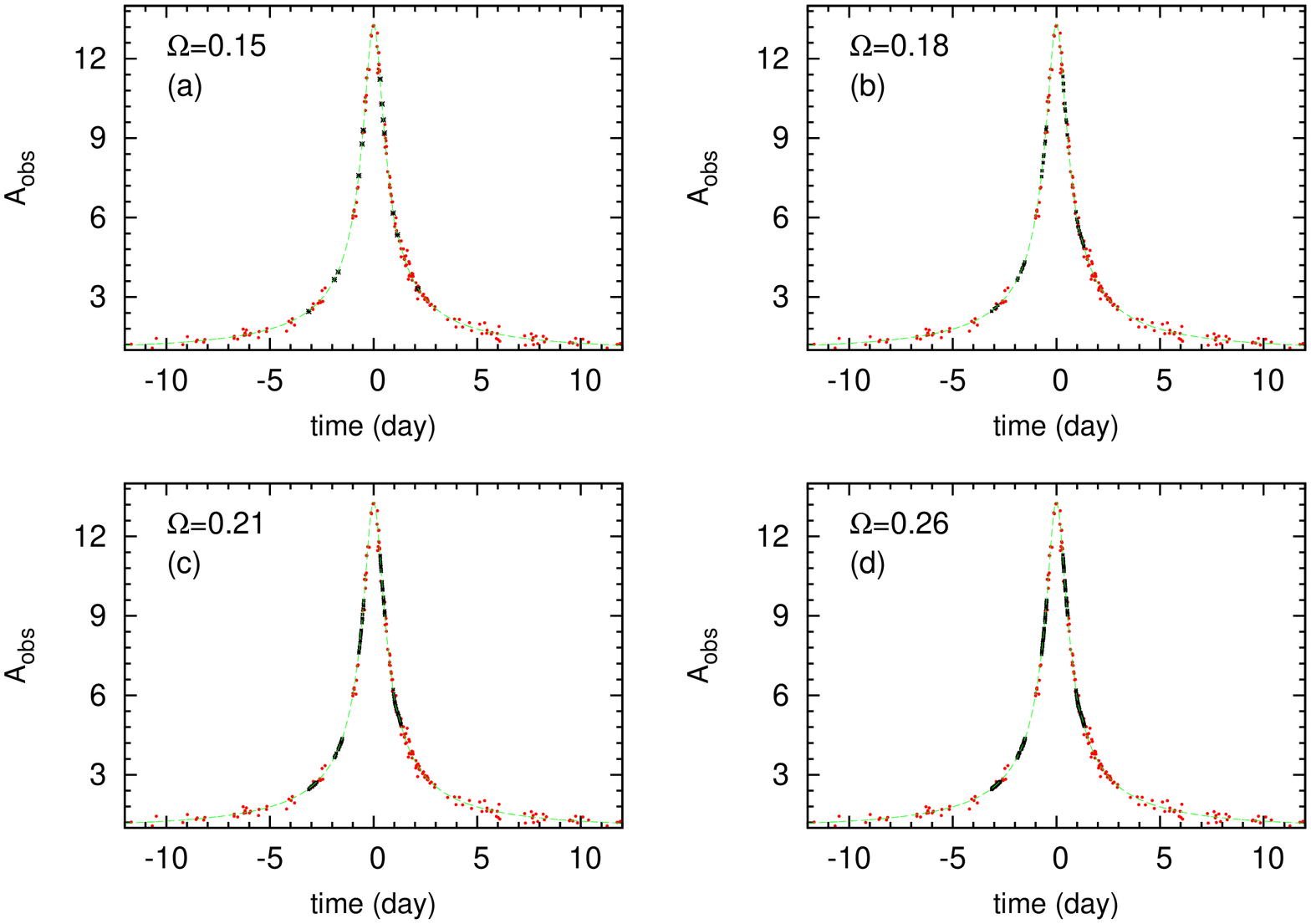,angle=270,width=0.8\textwidth,clip=0}
\caption{Imposing the factors $\mathrm{\kappa}$, all data points in
light curves shown in Figure (\ref{fig8}) are shifted to the least
amount of blending effect corresponding to the least amount of
$\mathrm{\Omega}$ (i.e. $\mathrm{\Omega=0.15~arcsec^{2}}$).
Therefore, all of these light curves have the same (and the least)
blending effects.} \label{fig8b}
\end{center}
\end{figure*}

The Danish telescope monitors the Galactic bulge for
$\mathrm{\sim6.5}$ hours on average each night during the observing
season. Also this telescope observes ongoing events during the night
with a probability of $\mathrm{\sim83}$ per cent \cite{Dominik10}
(i.e. no-observation might be due to weather condition, technical
failure and etc). We consider all of these observational limitations
in calculating cadences. To generate observational data points, the
magnification factors are shifted according to a Gaussian function
with the width of uncertainty in the magnification factor
$\delta_{A}$. This uncertainty is calculated according to the noise
$\mathrm{\sigma_{F}}$ and fractional systematic uncertainty for the
Danish telescope $\mathrm{\sigma_{s}/F_{obs}=0.3\% }$
\cite{Dominik10}. The overall amount of uncertainly is given by
$\delta_{A}=\sqrt{\sigma_{F}^{2}+\sigma_{s}^{2}}/F_{bg}$.

We consider different amounts of PSF area for the source star and
for each case we simulate the observations with the Lucky Imaging
camera on the Danish telescope. Figure (\ref{fig7}) shows a sample
of light curves with four different $\mathrm{\Omega}$s which depend
on the images we select from the lucky camera data. Here the
simulated data points hypothetically taken by the Lucky Imaging
camera on the Danish telescope are shown by black signs and data
points taken by other telescopes are shown with red circles. The
crucial role of the Lucky Imaging camera is to enhance small
perturbations in the light curve compare to the normal camera by
decreasing the blending effect.

Let us divide data in the light curve into two sets of being
observed by the Lucky camera with the Danish telescope and observed
by the normal cameras with the other telescopes. Then, in order to
fit the light curves with the theoretical model, we have different
blending parameters for each data set:
\begin{eqnarray}
\label{lc1}
A_{o}(t_{j})&=&b_{o} A_{\star}(t_{j})+(1-b_{o}) \\
A_{lc}(t_{j})&=&b_{lc} A_{\star}(t_{j})+(1-b_{lc}),
\label{lc2}
\end{eqnarray}
where $\mathrm{A_{\star}}$ is the magnification factor of the source
without considering blending effect, $\mathrm{A_{o}}$ is the
observed magnification factor with normal cameras and $\mathrm{b_o}$
is the corresponding blending parameter. For the Lucky camera,
$\mathrm{A_{lc}}$ is the observed magnification and
$\mathrm{b_{lc}}$ is the corresponding blending parameter. Having
less blending for the Lucky camera, means $\mathrm{b_{lc}}$ is
larger than $\mathrm{b_o}$. After fitting theoretical light curves
in equation (\ref{lc1}) and (\ref{lc2}) to the two sets of data
points, we calculate the ratio of larger blending parameter to the
smaller blending parameter, using the magnification factor from each
light curve
\begin{eqnarray}\label{blend12}
\alpha= \frac{A_{lc}(t_{j})-1}{A_{o}(t_{j})-1} =
\frac{b_{lc}}{b_{o}}.
\end{eqnarray}
Hence by imposing this factor, we can normalize the microlensing
data points by shifting low quality blending data points to the high
quality data points with the following transformation function:
\begin{eqnarray}\label{trans}
A'_{o}(t)=\alpha( A_{o}(t)-1)+1,
\end{eqnarray}
where $\mathrm{A'_{o}}$ is the new value of data points taken by
normal cameras. Figure (\ref{fig8}) shows this shifting for
different values of $\mathrm{\Omega}$s.
\begin{table*}
\begin{center}
\begin{tabular}{|c|c|c|c|c|c|c|c|}
&$\mathrm{\Omega (arcsec^2)}$ & $\mathrm{\bar{f}[\%]}$  & $\mathrm{\alpha_{n}\pm\Delta\alpha_{n}}$ &$\mathrm{\bar{t}_{exp}(s)}$  & $\mathrm{\bar{T}_{obs}(min)}$& $\mathrm{N}$& \\
\hline\hline
& $0.15$ & 1.7 & 3.9$\pm$0.073 & 203.2 & 45.4 & 206 &  \\
\hline
& $0.18$ & 5.0 & 3.5$\pm$0.069 & 163.4 & 44.9 & 236 &  \\
\hline
& $0.21$ & 12.0 & 3.2$\pm$0.061 & 162.4 & 21.0 & 291 &  \\
\hline
& $0.26$ & 23.9 & 2.9$\pm$0.056 & 159.8 & 10.1 & 404 &  \\
\hline
& $0.32$ & 43.0 & 2.4$\pm$0.049 & 155.3 & 5.6 & 579 & \\
\hline
& $0.40$ & 61.9 & 2.1$\pm$0.041 & 147.6 & 3.6 & 788 & \\
\hline
& $0.52$ & 93.4 & 1.8$\pm$0.036 & 144.3 & 2.5 & 1072 & \\
\hline
\end{tabular}
\end{center}
\caption{This table represents some properties of different light
curves corresponding to different amounts of the PSF area
$\mathrm{\Omega}$ whose amounts are brought in the first column. The
second and third columns show the average values of the percentage
of the selected images throughout the light curves
$\mathrm{\bar{f}[\%]}$ and $\mathrm{\alpha_{n}\pm \Delta
\alpha_{n}}$ corresponding to each light curve respectively. Two
next columns represent the averaged amounts of the effective
exposure time for data points hypothetically taken by Lucky Imaging
$\mathrm{\bar{t}_{exp}(s)}$ and the total observational time
$\mathrm{\bar{T}_{obs}(min)}$. $\mathrm{N}$ represents the number of
data points over each light curve.}\label{tabled}
\end{table*}

Shifting data points with the factor of $\alpha$, imposes
uncertainties in the data points due to the uncertainty in
$\mathrm{\alpha}$. Relative error in $\mathrm{\alpha}$ due to errors
in the blending parameters of each set of data points is given by:
\begin{eqnarray}
(\frac{\Delta\alpha}{\alpha})^{2}=(\frac{\Delta
b_{lc}}{b_{lc}})^{2}+(\frac{\Delta b_{o}}{b_{o}})^{2},
\label{errorb}
\end{eqnarray}
where $\mathrm{\Delta b_{lc}}$ and $\mathrm{\Delta b_{o}}$ are the
uncertainties in the blending parameters which can be obtained from
the maximum likelihood function of the best fit to the observed
data. The optimum amounts of blending parameters for the minimum
value of $\mathrm{\chi^2}$ obtains from $\mathrm{\partial
\chi^{2}/\partial b=0}$, results in
\begin{eqnarray}\label{blendeq}
b_{j}=\sum_i\frac{(A_{\star,i}-1)(A_{j,i}-1)}{\sigma^{2}}/\sum_{i}\frac{(A_{\star,i}-1)^{2}}{\sigma^{2}},
\end{eqnarray}
where $\mathrm{A_{j,i}}$ with  $\mathrm{j=lc,o}$ is the observed
value for the magnification and $\mathrm{A_{\star,i}}$ is intrinsic
value of the magnification factor at the $i$the time from the best
fit to the light curve. $\mathrm{\sigma}$ is the observational
photometric precision.
\begin{figure}
\begin{center}
\psfig{file=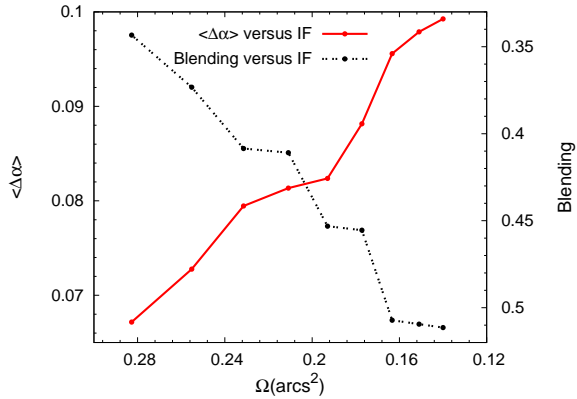,angle=0,width=8.cm,clip=} \caption{Result from
a Monte-Carlo simulation for the average value of
$\mathrm{\Delta\alpha}$ (red solid line) and the blending parameter
(black dotted line) versus the PSF area $\mathrm{\Omega}$.}
\label{fig9}
\end{center}
\end{figure}

Therefore, according to the uncertainty of $\mathrm{b}$ we can
obtain corresponding error of $\mathrm{\alpha}$ from equation
(\ref{errorb}). Figure (\ref{fig9}) shows the relation between the
error in $\mathrm{\alpha}$ and blending in terms of the PSF area for
an ensemble of events. While decreasing $\mathrm{\Omega}$ improves
the blending around the source star, it increases uncertainty in the
$\mathrm{\alpha}$ due to lack of enough data points in the light
curve. This effect also is shown in Table (\ref{tabled}). Finally,
the error bar corresponds to the magnification of the shifted data
due to the error in $\mathrm{\alpha}$ and error in the initial
magnification can be obtained from:
\begin{eqnarray}\label{cc}
(\frac{\Delta
A'_{o}}{A'_o})^{2}=(\frac{\Delta\alpha}{\alpha})^{2}+(\frac{\Delta
A_o}{A_o-1})^{2}.
\end{eqnarray}
Accordingly, the error bars of data points set at the baseline of
the microlensing light curves (i.e. $\mathrm{A_{o}\sim1}$) increase
significantly. Hence, the efficiency for detecting planetary signals
located around the baseline of the high-blended microlensing light
curves is not likely improved by the Lucky Imaging camera. These
planetary signals mostly happen while crossing the planetary caustic
curves close to the companion planets.

In Figure (\ref{fig8}), it is seen that the normalization of the
normal camera data points depends on PSF size that we adapted in the
lucky imaging observation. The smaller PSF, results in a higher
magnification factor. We may normalized all data points with
moderate PSF values to the smallest PSF that is generated only by
the best images from the lucky camera.
This procedure results in maximizing the data points in the light
curve. Let us define the best normalization factor with the best PSF
value by $\alpha_{max}$, then the shift factor between light curves
with different PSFs (i.e. $\alpha$s) can be defined as
$\mathrm{\kappa={\alpha_{max}}/{\alpha}}$. Figure (\ref{fig8b})
shows the shifted light curves with different $\mathrm{\Omega}$s and
$\mathrm{\alpha}$s with respect to the light curve with the smallest
$\mathrm{\Omega}$ (i.e. ${\mathrm{0.15~arcsec}}$). We note that
after shifting the data points, all light curves have normalized
with the same factor.


\begin{table*}
\begin{center}
\begin{tabular}{|c|c|c|c|c|c|c|c|c|c|c|c|c|}
&$\mathrm{\Omega (arcsec^2)}$ &$u_{0}$  & $t_{E}(day)$& $\theta^{\circ}$& $b$ & $d(R_{E})$ & $q(\times 10^{-5})$& $n$ & $\chi^{2}$  & $\Delta \chi^{2}$ & $\Delta \chi^{2}/n$& \\
\hline\hline
& $0.15$ & -0.04 & 12.0 & 25.0 & 0.51 & 0.9 & 2.0 & 2 & 71.1  & 0.5 & 0.25 & \\
&        & -0.04 & 12.0 & 25.0 & 0.51 &     &     &     & 70.6  &  &  & \\
\hline
& $0.18$  & -0.04 & 12.0 & 27.0 & 0.51 & 0.9 & 3.0 & 9 & 67.5 & 5.2 & 0.58 & \\
&         & -0.04 & 12.0 & 29.0 & 0.51 &     &     &     &  72.7 &  &  & \\
\hline
& $0.21$ &  -0.04 & 12.0 & 25.0 & 0.51 & 0.9 & 3.0 & 19 & 69.7 &  13.5 & 0.71 & \\
&        &  -0.04 & 12.0 & 29.0 & 0.51 &     &     &     &  83.2 &    &  & \\
\hline
& $\mathrm{0.26(LB)}$ &  -0.04 & 12.0 & 25.0 & 0.51 & 0.9 & 3.0 & 38 &  70.1  & 32.6 & 0.86 & \\
&        &   -0.04 & 12.0 & 29.0 & 0.51 &     &     &     &102.7  &  &  & \\
\hline
& $0.32$ &  -0.04 & 12.0 & 25.0 & 0.51 & 0.9 & 3.0 & 71 & 84.3 &  45.4 & 0.64 &\\
&        &  -0.04 & 12.0 & 29.0 & 0.51 &     &     &     & 129.7  &  &  & \\
\hline
& $0.40$ & -0.04 & 12.0 & 25.0 & 0.51 & 0.9 & 3.0 & 109    & 96.8  & 72.3 & 0.66 &\\
&        & -0.04 & 12.0 & 29.0 & 0.51     &   &     &      & 169.1  &  &  & \\
\hline
& $0.52$ &  -0.04 & 12.0 & 25.0 & 0.51 & 0.9 & 3.0 & 171 & 110.7 & 109.2 & 0.64 &\\
&        &   -0.04 & 12.0 & 29.0 & 0.51 &     &     &     &  219.9 &  &  & \\
\hline
& $\mathrm{1.02(LA)}$ &  -0.04 & 12.0 & 29.0 & 0.13 & 0.8 & 3.0 & 1   & 209.3  & 0.6 & 0.6 &\\
&        &               -0.04 & 12.0 & 27.0 & 0.13 &     &     &     & 209.9  &  &  & \\
\hline
\end{tabular}
\end{center}
\caption{This table shows the improvement in detectability of the
planetary signal using our strategy. The first column indicates the
amount of PSF area $(\Omega)$, corresponding to different light
curves some of them are represented in different panels of Figures
(\ref{fig7}), (\ref{fig8}) and (\ref{fig8b}). The light curve with
$\mathrm{\Omega=1.02~arcsec^2}$ is shown in Figure (\ref{fig6}) in
which we do not use any Lucky Imaging camera. The next columns
represent the best-fitted parameters from fitting different
planetary microlensing events to the mentioned light curves. The
theoretical parameters for making these light curves are
$\mathrm{u_{0}=-0.04}$, $\mathrm{t_{E}=12 days}$,
$\mathrm{\rho_{\star}=0.003}$, $\mathrm{\theta=25^{\circ}}$,
$\mathrm{t_{0}=0}$, $\mathrm{q=3\times10^{-5}}$ and
$\mathrm{d(R_{E})=0.9}$. We fixed $\mathrm{\rho_{\star}}$ and
$\mathrm{t_{0}}$ to their true values in the fitting process. We
note that the best fitted parameters as well as the minimum amounts
of $\chi^{2}$ from fitting simple Paczy\'nski microlensing models to
different light curves are brought in the bottom rows corresponding
to each light curve. $\mathrm{n}$ represents the number of data
points hypothetically taken by Lucky Imaging and cover the planetary
signal, i.e. located at the time interval of
$\mathrm{[0.9:1.25]~days}$. Two next columns show the minimum value
of $\mathrm{\chi^2}$ from fitting planetary (and simple Paczy\'nski)
microlensing events to the light curves as well as their
differences. The last column represents the normalized amounts of
$\Delta\chi^{2}$ to $\mathrm{n}$. The maximum amount of
$\mathrm{\Delta \chi^2/n}$ occurs for the light curve with
$\mathrm{\Omega=0.26~arcsec^{2}}$ and after shifting data points i.
e. $\mathrm{LB}$.}\label{tab1}
\end{table*}
\begin{figure}
\begin{center}
\psfig{file=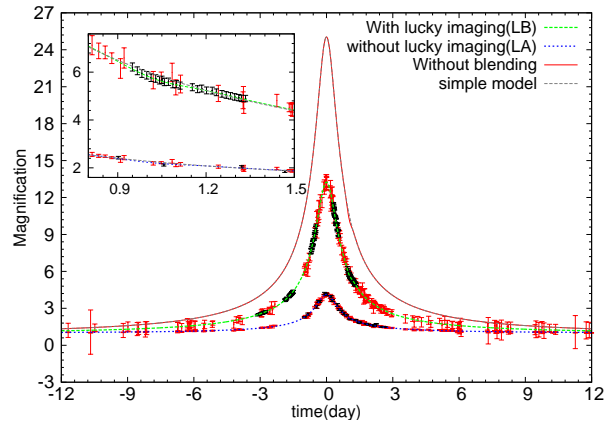,angle=-90,width=8.cm,clip=}\caption{This
figure summarizes the issues of this subsection and shows that (i)
the blending effect decreases the detectability of planetary signal
in the microlensing light curve and (ii) the Lucky Imaging camera
improves detectability of this signal by decreasing the blending
effect. The red solid line represents the sample planetary
microlensing event, studied in this subsection, without blending
effect. The gray dashed lines show the best-fitted simple
Paczy\'nski microlensing light curves with different blending
parameters. The blue dotted line shows that event with the blending
effect, $\mathrm{b=0.13}$. The simulated data points with the
corresponding error bars over this light curve are shown with the
black stars (taken by the Danish telescope) and the red circles
(taken by other telescopes hypothetically). This light curve is also
shown in Figure (\ref{fig6}), so-called LA. The green dashed line
and its data points represents that light curve resulted from this
assumption that the Danish telescope uses the Lucky Imaging camera
for detecting it, so-called LB. In this light curve the blending
effect improves up to the amount of $\mathrm{b=0.51}$, the error
bars of data points increase due to the shifting process while the
planetary signal becomes more detectable. The planetary signals of
these light curves are enlarged in the inset of the figure.}
\label{fig11}
\end{center}
\end{figure}

We summarize the properties of the generated light curves
corresponding to different amounts of the PSF area in the table
(\ref{tabled}). In this table, the second and third columns show the
average values of the percentage of the selected images throughout
the light curves $\mathrm{\bar{f}[\%]}$ and $\mathrm{\alpha_{n}\pm
\Delta \alpha_{n}}$ corresponding to each light curve respectively.
Two next columns represent the averaged amounts of the effective
exposure time for data hypothetically taken by Lucky Imaging
$\mathrm{\bar{t}_{exp}(s)}$ and the total observational time
$\mathrm{\bar{T}_{obs}(min)}$. $\mathrm{N}$ represents the number of
data points over each light curve. Noting that the effective
exposure time depends on the PSF area and decreases by increasing
it. Because we fix the photometric error bars throughout the light
curves. The total observational time depends on the fraction of the
selected images and increases by decreasing this factor.

In this part we examine the best value for the PSF size that we can
choose from the lucky imaging observation. Our criterion is to
maximize the signal to noise ratio of the planetary signal in the
binary lensing with minimizing the  $\mathrm{\chi^2}$ value from
fitting to the model. We subtract  $\mathrm{\chi^2}$ from the binary
lensing and simple Paczy\'nski microlensing  (i.e.
$\mathrm{\Delta\chi^{2}= \chi_0^{2} - \chi_{pl}^{2}}$) where large
$\mathrm{\Delta\chi^{2}}$ indicates a better signal from the
lensing. Table (\ref{tab1}) shows the best-fitted parameters from
fitting different planetary microlensing events as well as
$\mathrm{\Delta \chi^{2}}$ in terms of various values of
$\mathrm{\Omega}$ which corresponds to different light curves. This
table shows that using a Lucky Imaging camera and shifting data
points (a) improves the detectability of a planetary signal
according to the criterion of $\mathrm{\Delta \chi^{2}}$ and (b)
approaches the best-fitted parameters to their true values.
According to this table, the light curve with the PSF area
$\mathrm{\Omega=0.26~arcsec^{2}}$ with the largest number of data
points has the maximum amount of $\mathrm{\Delta \chi^{2}}$ and its
best-fitted parameters coincide to their real values. Hence, this
strategy benefits from the two important factors of (i) high
resolution images at the first round of analysis of the Lucky camera
by selecting a low percentage of good images (ii) shifting all the
data points of larger PSF area with the corresponding blending
factor in order to decrease the cadence of the data points in the
light curve. We note that the photometric accuracy for all data
points taken by the Lucky camera with different $\mathrm{\Omega}$s
is fixed.

To clarify how much this strategy improves the quality of the
planetary signal in microlensing light curves, we compare
reconstructed light curve which has the highest amount of
$\mathrm{\Delta\chi^{2}}$, i.e. that light curve corresponding to
$\mathrm{\Omega=0.26~arcsec^2}$ after shifting to the least blending
effect, with the original light curve taken from the normal cameras
(shown in Figure \ref{fig6}). We called the former light curve
before shifting the other data points by LB and the later one after
shifting with LA. Figure (\ref{fig11}) shows these light curves with
simulated data points with their error bars, the theoretical light
curve with no blending effect (red solid curve) and the best-fitted
simple Paczy\'nski microlensing model (gray dashed curves). The
planetary signals of these light curves is zoomed in the inset of
the figure. The Lucky Imaging camera decreases the blending effect,
but increases the photometric uncertainties a bit (due to shifting
process) and improves the detectability of the planetary signal (see
Table \ref{tab1}). Here, the ratio of the blending parameters of
these two light curves is $\mathrm{\alpha_{1}=3.9}$ which means that
the blending effect in the LB light curve decreases by a factor of
$\mathrm{1/\alpha_{1}}$.

\begin{figure*}
\begin{center}
\psfig{file=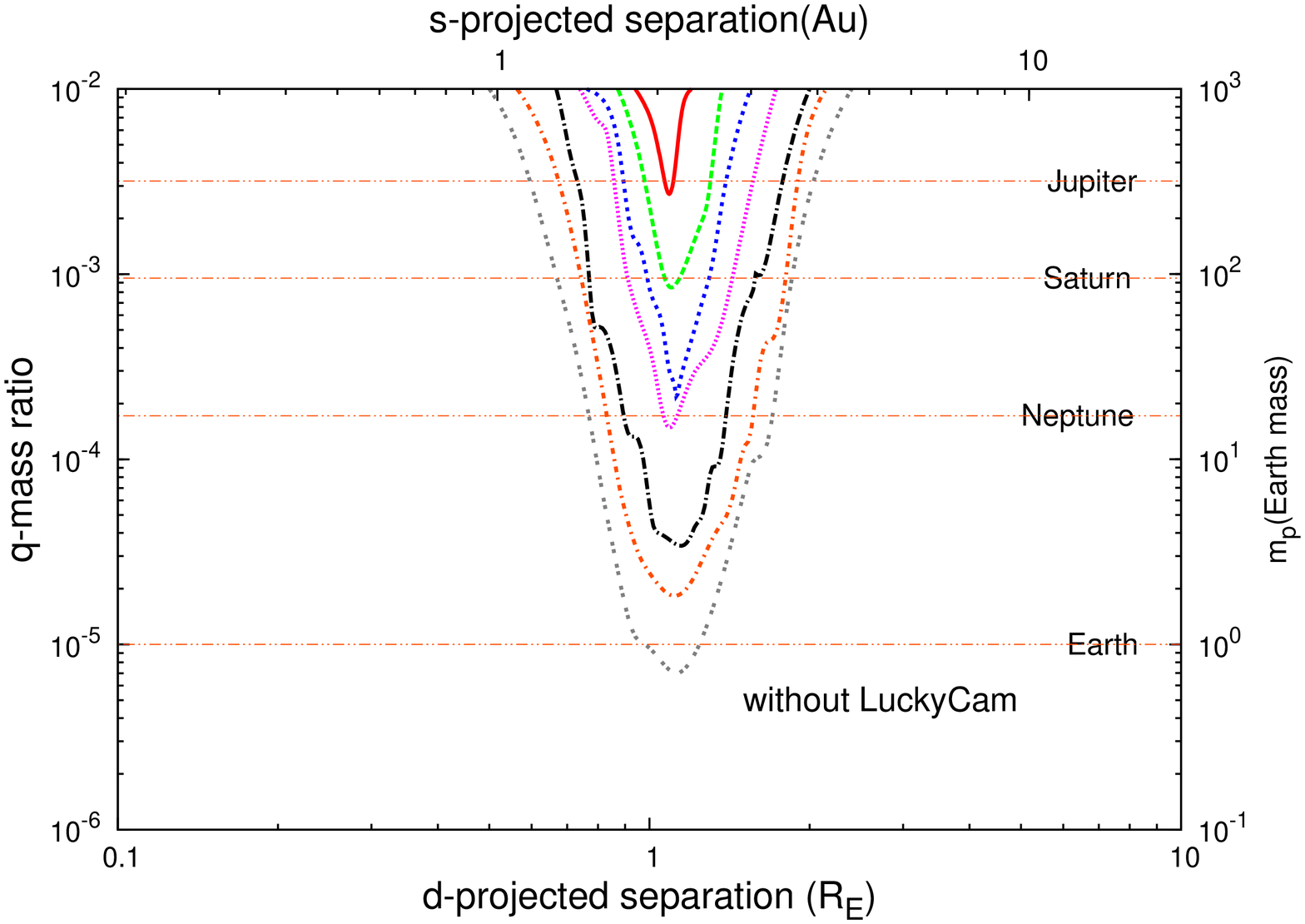,angle=270,width=0.495\textwidth,clip=0}\label{fig2b}
\psfig{file=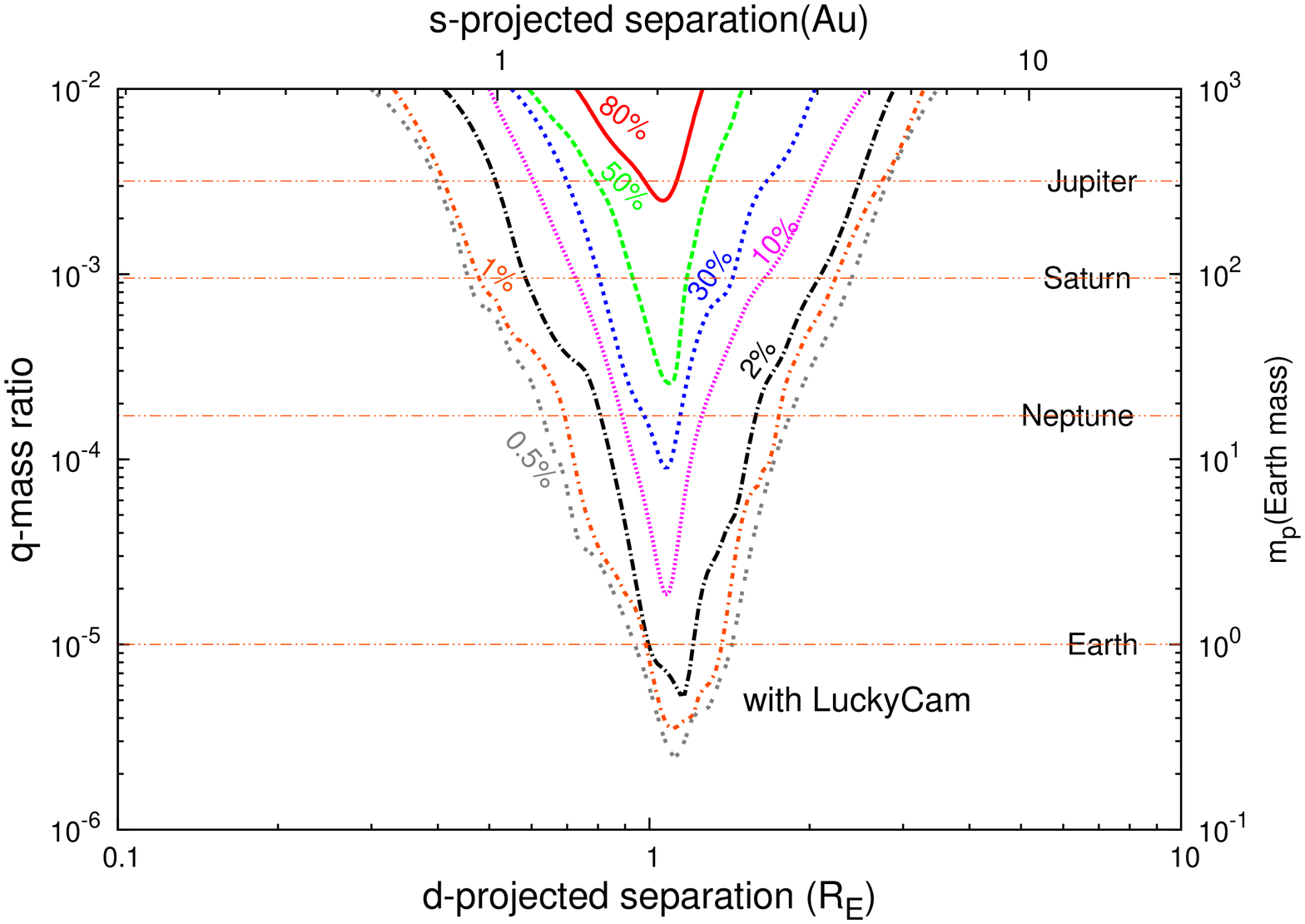,angle=270,width=0.495\textwidth
,clip=0}\label{fig2c} \caption{The contour lines of planet detection
efficiency for some of blended events from OGLE 2012 microlensing
events. In the left panel we use the normal camera on the Danish
telescope and in the right panel, we assume the Danish telescope is
occupied with a Lucky camera and the introduced strategy is used to
make light curves. For converting $\mathrm{\varepsilon(d,q)}$ to
$\mathrm{\varepsilon(s,m_{p})}$ we use the typical amounts for the
the mass of the primary lens $\mathrm{M_{l}=0.3 M_{\odot}}$, the
distances of the lens and source from the observer
$\mathrm{D_{s}=8.5~kpc}$ and $\mathrm{D_{l}=6.5~kpc}$.}\label{fig12}
\end{center}
\end{figure*}

\section{Planet detection efficiency}
\label{detection} We continue our study to estimate the fraction of
planets that can be detected by a typical microlensing event if we
use the Lucky camera according to the strategy discussed in the
previous section. We adopt the formalism introduced by Gaudi \&
Sackett (2000a,b) for detection efficiency calculation of real
microlensing events with the blending effect.

We consider some of OGLE microlensing events with non-zero blending
effects. We calculate the planet detection efficiency of every event
and finally overlap them. For each event, we pick up its parameters
from fitting process: i.e. the Einstein crossing time
$\mathrm{t_{E}}$, the time of the closest approach $\mathrm{t_{0}}$,
the lens impact parameter $\mathrm{u_{0}}$, the blending parameter
$\mathrm{b}$ and the baseline flux $\mathrm{F_{bg}}$. The apparent
source magnitude, $\mathrm{m_{\star}}$, can be obtained according to
the blending parameter and the baseline flux. To indicate the source
star radius of each microlensing event, we first specify the type of
the source according to its apparent magnitude. Indeed, amongst the
observed microlensing targets, there is a bimodal distribution for
the source star magnitude, corresponding to bright giants on one
side and highly-magnified faint main-sequence stars on the other
side \cite{cassan}. The transition magnitude between these two types
of giant and main-sequence stars is $17$ mag. We consider
$\mathrm{R_{\star}=10 R_{\odot}}$ for the former and
$\mathrm{R_{\star}= R_{\odot}}$ for the latter, where
$\mathrm{R_{\odot}}$ is the sun radius. For indicating the projected
source radius on the lens plane $\mathrm{\rho_{\star}}$, we need to
determine the Einstein radius and the lens-source relative distances
from observer $\mathrm{x=D_{l}/D_{s}}$.

Estimations of these two parameters as well as indication of the
lens mass are done according to the probability density of these
parameters versus the Einstein crossing time which is plotted in
Figure (5) of Dominik (2006). Two last parameters are times when the
first $\mathrm{(t_{min})}$ and the last $\mathrm{(t_{max})}$ data
points are taken by the Lucky Imaging camera for each light curves.
Since a trigger is necessary to start observations it makes sense to
continue following microlensing events further down after the peak,
we adopt to start taking data points at $\mathrm{A_{obs}=1.5}$ and
to stop at $\mathrm{A_{obs}=1.06}$.

For each event, the parameters of the planetary system, $\mathrm{q}$
and $\mathrm{d}$, change uniformly in the logarithmic scale over the
ranges of $\mathrm{q\in[10^{-6},10^{-2}]}$ and
$\mathrm{d\in[0.1,10]}$ with steps $\mathrm{\Delta \log q= 0.0625}$
and $\mathrm{\Delta \log d= 0.03125}$, respectively. The angle
between the trajectory of the source with respect to the binary axis
changes in the range of $\mathrm{\theta\in[0,360^{\circ}]}$ with the
steps of $\mathrm{\Delta \theta=0.36^{\circ}}$.

For each configuration of lenses in terms of $\mathrm{q}$ and
$\mathrm{d}$, the planet detection efficiency
$\mathrm{\varepsilon(q,d)}$ is obtained for all the possible source
trajectories $\mathrm{(0<\theta<2\pi)}$ which is given by:
\begin{eqnarray}
\varepsilon(q,d)= \frac{1}{2\pi}\int_{0}^{2\pi}\Theta(q,d) d\theta.
\end{eqnarray}
where $\mathrm{\Theta}$ is a step function and is calculated for
each simulated planetary microlensing event corresponding to given
amounts of $\mathrm{(d,q,\theta)}$. We apply this function when in
the related light curve four consecutive data points are located
above (or below) the Paczy\'nski light curve with more than
$\mathrm{2\sigma}$ deviation. These points should be in the same
side with respect to the Paczy\'nski light curve. Otherwise, this
function is set to be zero.

Having performed the simulation, in Figure (\ref{fig12}) we compare
the efficiency function of the planet detection using the standard
camera in the Danish telescope (left panel) and the Lucky Imaging
camera (right panel). For converting $\mathrm{\varepsilon(d,q)}$ to
$\mathrm{\varepsilon(s,m_{p})}$ we use typical values for the the
mass of the primary lens $\mathrm{M_{l}=0.3 M_{\odot}}$, the
distances of the source and the lens from the observer
$\mathrm{D_{s}=8.5~kpc}$ and $\mathrm{D_{l}=6.5~kpc}$, where
$\mathrm{s}$ is the projected separation in the Astronomical Unit
(AU) and $\mathrm{m_{p}}$ is the planet mass. Comparing these two
figures shows that the Lucky Imaging camera improves the planet
detection efficiency in a crowded Field. Noting that we consider
only microlensing events with non-zero blending effect. For example
a Neptune-mass planet in the resonance regime can be detected with
$30$ per cent chance with the Lucky camera compare to the
normal camera with $10$ per cent probability. 
Also, we obtain the detection efficiency of an Earth-mass planet in
the resonance regime is more than $\mathrm{\varepsilon \sim 2}$ per
cent with the Lucky camera while with the normal camera this
detection efficiency is about $\mathrm{0.5\%}$ in a crowded-field.
If the abundance of Earth-mass planets is high (which is probable
according to Cassan et al. 2012), a considerable number of
Earth-mass planets in the resonance regime would be expected for
detection with this method.


\section{Conclusions}
\label{con} In this work we have investigated the advantage of using
Lucky Imaging for exoplanet detection via the microlensing method.
The Lucky Imaging method is a technique where cameras take series of
images faster than the time scale of atmospheric turbulence. The
selected best images where wave-fronts are not distorted are
re-centered and combined to improve the final image.

We have simulated microlensing events observed with the Lucky camera
on Danish $1.54$-m telescope accompanied by normal CCD cameras with
other telescopes and in this regard introduced an observational
strategy. We propose to fix the PSF area of the source star for a
fixed photometric precision for all data points in a microlensing
event. For a fixed PSF area in the Lucky camera the blending of the
source star is fixed. We have two sets of data points with different
blending parameters one for the normal camera and the other one with
the Lucky camera. We propose shifting the low quality blending data
points to the high quality data points by calculating the ratio of
their blending parameters (i.e. $\mathrm{\alpha}$) in equation
(\ref{blend12}). Re-positioning some of data points over a
microlensing light curve imposes an uncertainty for the shifted data
points in which these uncertainties decrease by the PSF area. On the
other hand we get fewer data points throughout the light curve as we
decrease the PSF area (see Table \ref{tabled}).

In order to address that problem, we propose to produce data points
from the Lucky camera with the least PSF area. Since these Lucky
images are stored in our database, we repeated analyzing Lucky
camera data with larger $\mathrm{\Omega}$s, resulted in increasing
the data points in the light curve. From the $\mathrm{\alpha}$
parameter in each data set of the Lucky camera, we shifted the
corresponding data points to that of the highest $\mathrm{\alpha}$.
Finally, we produced a light curve with the minimum blending effect
and highest number of data points. In our analysis we have the best
light curve according to the maximum value of $\mathrm{\Delta
\chi^{2}}$, representing the difference between $\mathrm{\chi^2}$
from fitting a simple microlensing light curve and a planetary
microlensing light curve.

We have also studied the improvement of the planetary detection
efficiency using the Lucky Imaging camera. In this regard, we have
microlensing events of OGLE 2012 data with nonzero blending effect,
calculated their planetary detection efficiencies and overlapped
their detection efficiencies. We obtained the detecting efficiency
of an Earth-mass planet with the Lucky camera in the resonance
regime increases up to $\mathrm{\sim 2\%}$ while this efficiency
with the normal cameras is about $\mathrm{0.5\%}$ in a crowded
field. If the abundance of these planets is high (which is probable
according to Cassan et al. 2012), the number of detectable
Earth-mass planets whose positions are projected in the resonance
regime will be four times more with using this method.

\textbf{Acknowledgment} Works by SS and SR were supported by the
grant $\mathrm{94017434}$ from the Iran National Science Foundation
(INSF). We thank the referee for useful comments and suggestions
which certainly improved the manuscript.


\end{document}